# Direct optical excitation of an NV center via a nanofiber Bragg-cavity: A theoretical simulation

TOSHIYUKI TASHIMA[1*], HIDEAKI TAKASHIMA[1], AND SHIGEKI TAKEUCHI[1**]

[1]*Department of Electronic Science and Engineering, Kyoto University, 615-8510 Kyoto, Japan*

*tashima.toshiyuki.5e@kyoto-u.ac.jp

**takeuchi@kuee.kyoto-u.ac.jp

**Direct optical excitation of a nitrogen-vacancy (NV) center in nanodiamond by light via a nanofiber is of interest for all-fiber-integrated quantum applications. However, the background light induced by the excitation light via the nanofiber is problematic as it has the same optical wavelength as the emission light from the NV center. In this letter, we propose using a nanofiber Bragg cavity to address this problem. We numerically simulate and estimate the electric field of a nanodiamond induced by excitation light applied from an objective lens on a confocal microscope system, a nanofiber, and nanofiber Bragg-cavities (NFBCs). We estimate that by using a nanofiber, the optical excitation intensity can be decreased by roughly a factor of 10 compared to using an objective lens, while for an NFBC with a grating number of 240 (120 for one side) on a nanofiber the optical excitation intensity can be significantly decreased by roughly a factor of 100. Therefore, it is expected that the background light inside a nanofiber can be significantly suppressed.** © 2015 Optical Society of America

http://dx.doi.org/10.1364/OL.99.099999

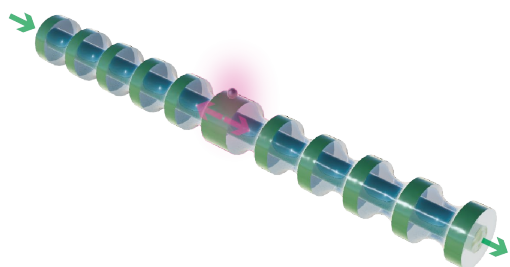

Fig. 1. Overview of the optical excitation of a nanodiamond by excitation light via a nanofiber Bragg-cavity (NFBC). The nanodiamond is located at the center of the NFBC. The green arrows are the excitation laser, and the red arrows are the emission light from an NV center in a nanodiamond.

All-fiber-integrated quantum applications are highly promising candidates for quantum information processing and communications. To realize such devices, the coupling of nanoparticles, such as nanodiamond, to nanofibers has been experimentally investigated [1–23]. In particular, nitrogen-vacancy (NV) centers in nanodiamonds [24] coupled to optical nanofibers is the most promising candidate for realizing fiber-based quantum applications. To increase the optical detection efficiency of emission light from NV centers in nanodiamonds coupled to nanofibers, many theoretical and experimental demonstrations have made undertaken [1–4, 6, 8–11, 13, 14, 17–20], such as the manipulation of nanodiamond on nanofibers by atomic force microscopy (AFM) [10], and nanofiber coupling to diamond micro waveguides [19]. Also, optically detected magnetic resonance (ODMR) for NV centers in nanodiamonds on nanofibers has been experimentally demonstrated [25–27].

The optical excitation of nanodiamonds can be achieved using excitation light via an objective lens on a confocal microscope system. To realize a genuine all-fiber system, however, the optical excitation should be performed directly via the nanofiber. However, the emission light from an NV center via nanofiber is hidden by background light induced by the excitation light from the nanofiber, which has the same optical wavelength [26], greatly decreasing the signal to noise ratio of the detection efficiency of the emission light from the NV center. To address this problem, a lock-in amplifier was necessary to isolate the signal via the nanofiber from the background light in ODMR measurements [26]. Thus, it is important to concentrate on an optical excitation that is highly enhanced at only a certain point of an NV center, while suppressing background light.

Here, we focus on a nanofiber Bragg cavity (NFBC) [15, 16], which is a series of Bragg cavities built into a nanofiber, as shown in Fig. 1. Using an NFBC will allow a nanodiamond at a point on the nanofiber to be strongly excited with weak input light. Thus, we expect that we can achieve a high optical excitation while suppressing the background light inside the nanofiber.

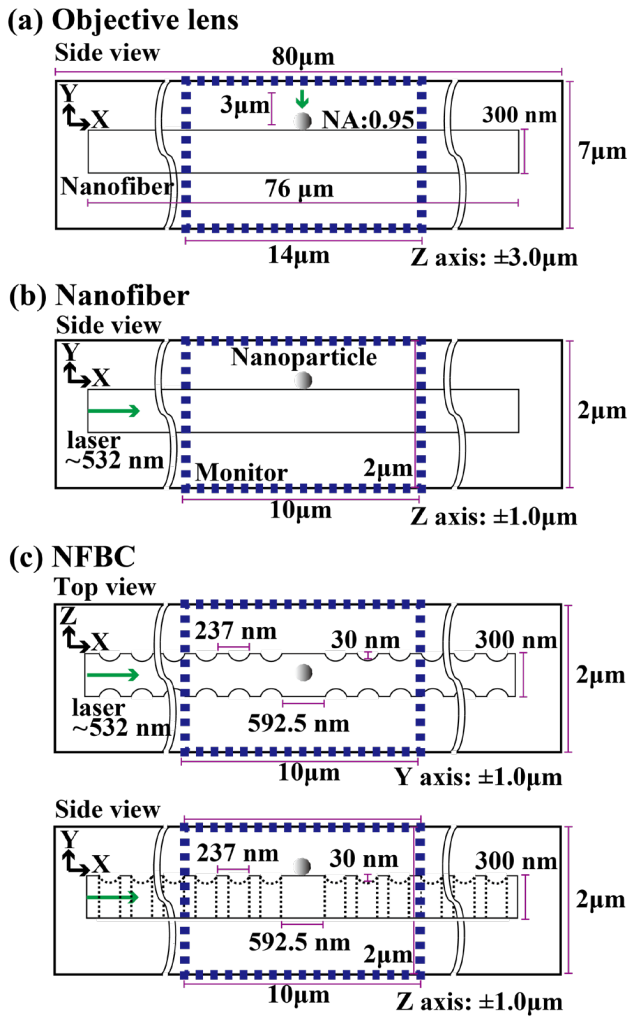

Fig. 2. Conditions of simulation of electric field by excitation light via an objective lens (a), nanofiber (b), and NFBC (d). For (a), the light source is placed 3 μm from the center of the nanodiamond. For (b) and (c), the light source is placed at the end of a nanofiber. The blue box is a monitor for observing the electric field around a nanodiamond on an NFBC.

3D FDTD simulations (FDTD Solutions, Lumerical) were performed to numerically simulate and estimate the electric field of a nanodiamond on a nanofiber. For the structure of the nanodiamond, a polyhedron structure with a refractive index of 2.417 was used. The number of polyhedrons was more than 100 and the size of the nanodiamond was 25 nm. The diameter and length of the nanofiber were 300 nm and 76 $\mu$m, respectively. The material of the nanofiber was $SiO_2$ with a refractive index of ~ 1.46.

Figure 2(a) shows the geometry of our calculation model for excitation of the nanodiamond via an objective lens. The calculation region ($L \times W \times Z$) was set to $\pm 3.5_\mu$m $\times \pm 3.5_\mu$m $\times \pm 3_\mu$m. A perfectly matched layer (PML) was used as the boundary condition. An automatic nonuniform mesh was used for the calculation region and a uniform mesh with a size of 2.5 nm was overlaid around the nanodiamond. The time step and simulation time were set to 0.03 fs and 50 ps, respectively. As the light source to excite the nanodiamond, we used a Gaussian source with a numerical aperture (N. A.) of 0.95 and a beam diameter of 5.5 $\mu$m. This source was placed 3 $\mu$m from the center of the nanodiamond. The wavelength of the light was 532 nm and the power was 1 mW. The input power was the same for all the simulations.

Figure 2(b) shows the calculation geometry for excitation of the nanodiamond by laser light through the optical nanofiber. The calculation region with a PML boundary condition was 80μm ×±1μm ×±1μm. The region where the electric field was monitored was 10μm ×±1μm ×±1μm. The light source was placed at the end of the nanofiber. The injected light had an electric field distribution of the fundamental mode of the nanofiber with a wavelength of 532 nm.

Figure 2(c) shows the calculation geometry for excitation of the nanodiamond by the NFBC. The calculation region and the monitoring region were the same as for the case for the nanofiber. As an approximation of the grating of an actual NFBC, we used cylindrical grooves on the top and both sides of the nanofiber[15, 16]. The depth of the grooves was 30 nm and the period of the grating was 273 nm. These values gave an NFBC that operated at a wavelength of around 532 nm. At the position of the nanodiamond, there was a gap in the grating of 592.5 nm, which divided the grating into two "sides", one before the nanodiamond and one after. The grating number on one side of the nanofiber was changed from 20 to 120 in steps of 20. The light source of the NFBC was placed at the end of the nanofiber. The transmitted power was monitored to calculate the transmission spectra at the other end of the nanofiber.

Figure 3(a) shows the cross-section (X–Y plane) of the electric field distribution around the nanodiamond when it was excited using the objective lens. The intensity of the electric field is strong around the nanofiber. The inset of Fig. 3(a) shows a magnified image around the nanodiamond, showing that the intensity drops inside the nanodiamond. Figure 3(b) shows the cut line at X = 0 μm of the electric field distribution along the white dotted line in Fig. 3(a). The peak intensity of the electric field is 1.7 MV/m near the surface of the nanodiamond (Y = 0.15 μm). The intensity of the electric field at the center of the nanodiamond is 0.46 MV/m because of Rayleigh scattering with the fact that the size of the nanodiamond was less than the wavelength of the light source [28]. When there was no nanodiamond in the calculation region, the intensity of the electric field at the position of the nanodiamond was 1.8 MV/m. That is, the field intensity is decreased by a factor of 0.26.

In Fig. 3(c), we show the electric field for excitation of the nanodiamond by the nanofiber. The electric field is localized at the surface of the nanofiber and at the center of the nanofiber. This is due to the fact that the diameter of the nanofiber was smaller than the wavelength of the propagation light[11]. The electric field inside the nanodiamond drops, as shown in the inset of Fig. 3(c). Figure 3(d) shows the cut line of the electric field at X = 0 μm. The electric field at the center in the nanodiamond is 1.2 MV/m, which is about 2.6 times larger than that for the objective lens. Hence the same electric field intensity can be realized with a laser power of 0.15 times the power of the objective lens.

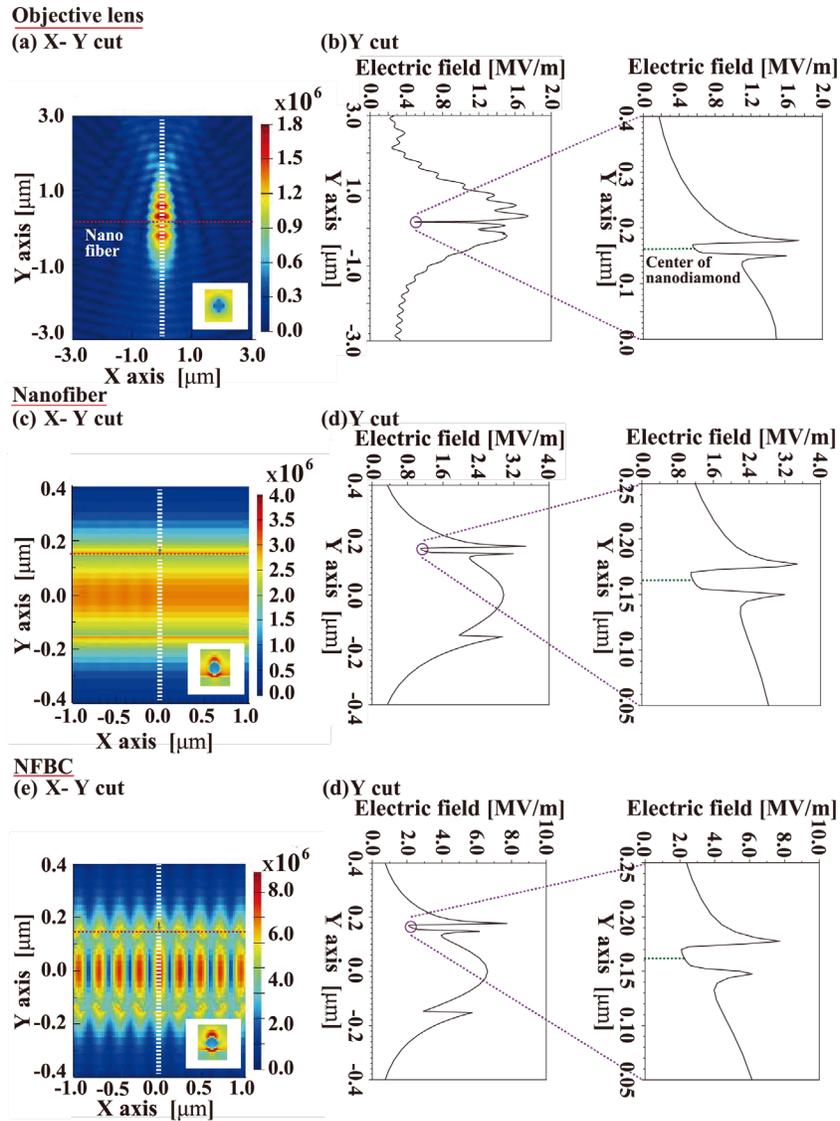

Fig. 3. Simulation results of an electric field induced by light via an objective lens, a nanofiber, and an NFBC. (a), (b): X–Y cut and Y cut of the electric field induced by light via an objective lens. (c), (d): X–Y cut and Y cut of the electric field induced by light via a nanofiber. (e), (f): X–Y cut and Y cut of the electric field induced by light via an NFBC (grating number of 80 on one side). The images at the bottom right in (a), (c), and (e) are magnified images of the electric field around the nanodiamond.

To realize direct optical excitation of the nanodiamond with an even smaller excitation power, we considered the distribution of the electric field using an NFBC with a grating number of 80 on one side. Figure 3(e) shows the electric field distribution. Figure 3(f) shows the cut line of the electric field at X = 0 $\mu$m. The electric field at the center of the nanodiamond is 2.6 MV/m, which is an increase of 5.7 times over the objective lens case. This means that an optical power of 31 $\mu$W is sufficient to generate the same electric field as generated by an objective lens with an optical power of 1 mW, i.e., the optical excitation intensity can be decreased by ~3/100 times by using this NFBC. The power when using the NFBC also decreased in comparison with that of the nanofiber. It is thus expected that background photons induced from excitation light inside the nanofiber can be significantly suppressed.

We also calculated the transmission spectrum of the NFBC to confirm the loss due to the grating structure with the fluorescence wavelength of the NV centers. Figure 4 shows the calculation result in the range 500 nm to 700 nm for a grating number on one side of the NFBC of 120 for considering following simulation. The fluorescence wavelength of the NV centers is roughly from 600 nm to 700 nm and the transmission for these regions is over 0.8. It is noted that the shape peak around 532nm is the resonant peak of an NFBC. This result shows that the fluorescence from the NV centers was not affected significantly by the NFBC with a resonant wavelength of 532 nm.

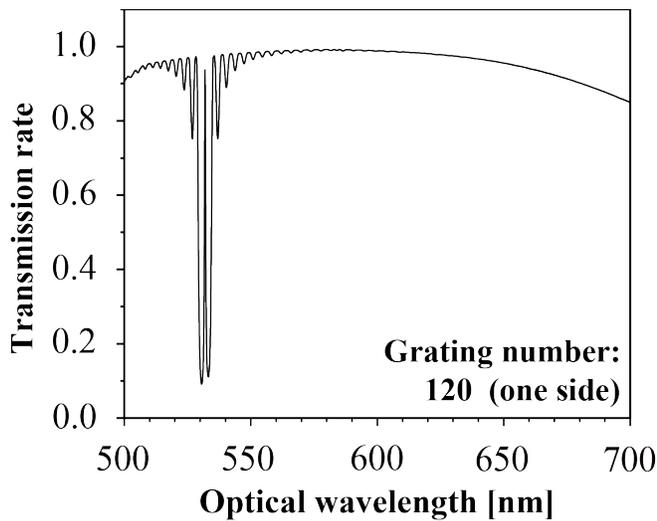

Fig. 4. Transmission spectrum of an NFBC with a grating number of 120 for one side.

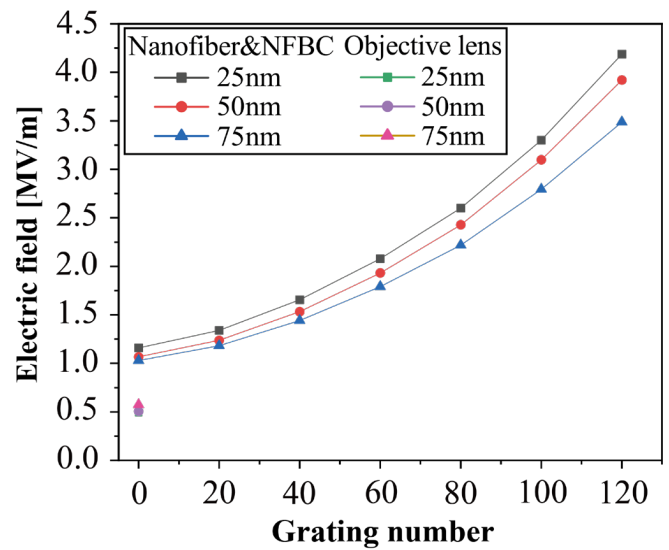

Fig. 5. Simulation results of the electric field at the center of a nanodiamond changing the grating number of the NFBC and the size of the nanodiamond. The electric fields with an objective lens are also shown for each size of the nanodiamond. Note that this grating number refers to one side of the NFBC.

Finally, we calculate the dependence of the electric field inside the nanodiamond on the grating number of the NFBC. The grating number of one side of the NFBC was changed from 0 to 120 in steps of 20. The sizes of the nanodiamonds used were 25 nm, 50 nm, and 75 nm. The results are shown in Fig. 5. As a reference, we also show the electric fields for optical excitation via an objective lens in the figure. These electric fields for the objective lens are around 0.5 MV/m for each nanodiamond size. This shows that the dependence on the size of nanodiamond was low because the input light was scattered for the whole nanodiamond with a size less than the wavelength.

On the other hand, for the NFBCs, when the size of a nanodiamond increased, the electric field decreased according to the size of the scattering cross section of the nanodiamond. Also, when the grating number increased, all three nanodiamond sizes show an increase in the electric field. Consider for example a grating number of 120 for one side. The electric fields for nanodiamond sizes of 25 nm, 50 nm, and 75 nm are 4.2 MV/m, 3.9 MV/m, and 3.5 MV/m, respectively. The 50- and 75-nm nanodiamonds have electric fields of, respectively, 0.9 times and 0.8 times compared to that of the 25-nm nanodiamond. Thus, the decrease of the electric field due to nanodiamond size was small. The electric field of a nanodiamond of 25 nm and a grating number of 120 of 4.2 MV/m is 9.1 times as large as the value of 0.46 MV/m for an objective lens. This means that an optical power of 12 $\mu$W ($1.2 \times 10^{-2}$ times) is sufficient to generate the same electric field as an objective lens with an optical power of 1 mW. Namely, the optical excitation intensity using a nanofiber can be decreased by ~1/100 times.

In conclusion, we have numerically simulated the electric field in a nanodiamond driven by light from an objective lens, a nanofiber, and NFBCs. We find that a nanofiber can generate an electric field for an optical excitation intensity ~1/10 times that required for an objective lens. The optical intensity required for an NFBC can be ~1/100 times lower. Note that we performed simulations for NFBCs with grating numbers up to 120, but an NFBC has been experimentally fabricated with a grating number of 160 for one side [29]. While experiments to date have been performed for excitation light from an objective lens on a confocal microscope system, our results indicate that direct optical excitation by light from an NFBC can be used with a huge decrease in optical power. Thus, our results open up a new method for genuine fiber-integrated quantum applications.

**FUNDING INFORMATION**

Part of this work is supported by a Kakenhi Grant-in-Aid (No. 26220712) from the Japan Society for the Promotion of Science (JSPS), MEXT Q-LEAP, and CREST program of the Japan Science and Technology Agency (JST) (JPMJCR1674).

**ACKNOWLEDGEMENTS**

We thank Y. Iwabata, H. Maruya, and K. Fukushige for the valuable comments on our work.